# Dry granular flows: rheological measurements of the $\mu(I)$-rheology

A. Fall[1], G. Ovarlez[1,2], D. Hautemayou[1], C. Mézière[1], J.-N. Roux[1] and F. Chevoir[1]

[1] *Université Paris Est, Laboratoire Navier* (CNRS, IFSTTAR, Ecole des Ponts ParisTech), *Champs-sur-Marne, France*

[2] *CNRS, LOF, UMR 5258, F-33600 Pessac, France*

Dated 27 April 2015

## Synopsis

Granular materials do not always flow homogeneously like fluids when submitted to external stress, but often form rigid regions that are separated by narrow shear bands where the material yields and flows. This shear localization impacts their apparent rheology, which makes it difficult to infer a constitutive behaviour from conventional rheometric measurements. Moreover, they present a dilatant behaviour, which makes their study in classical fixed-volume geometries difficult. These features led numerous groups to perform extensive studies with inclined plane flows, which were of crucial importance for the development and the validation of the $\mu(I)-$ rheology. Our aim is to develop a method to characterize granular materials with rheometrical tools. Using rheometry measurements in an annular shear cell, dense granular flows of 0.5 mm spherical and monodisperse beads are studied. A focus is placed on the comparison between the present results and the $\mu(I)-$ rheology. From steady state measurements of the torque and the gap under imposed shear rate $\dot{\gamma}$ and normal force $F_N$, we define an inertial number $I$. We show that, at low $I$ (small $\dot{\gamma}$ and/or large $F_N$), the flow goes to a quasi-static limit, and the response in terms of dimensionless stress or internal friction coefficient $-\mu-$ and solid concentration $-\phi-$ profiles is independent of the inertial number. Upon increasing $I$ (large $\dot{\gamma}$ and/or small $F_N$), dilation occurs and $\phi$ decreases while $\mu$ increases. The observed variations are in good agreement with previous observations of the literature (Jop et al. 2006; Hatano 2007). These results show that the constitutive equations $\mu(I)$ and $\phi(I)$ of granular materials can be measured with a rheometer.

# I. Introduction

Granular matter shows both solid and fluid behavior. Of interest in many industrial processes and in geophysics, granular flows are the focus of very active researches (Duran 2000). These materials are very sensitive to various parameters: geometry of the flow, wall roughness, flow rate, shape and size distribution of the grains, and coupling with the interstitial fluid (Andreotti et al. 2013). Due to their macroscopic size, the interactions between the grains are dissipative (friction and inelastic collisions); the energy lost is then transferred to internal degrees of freedom. The lack of Brownian motion and the dissipative interactions, make the granular material an intrinsically nonequilibrium system.

In the dry case – *without interstitial fluid* –, the rheology is solely governed by momentum transfer and energy dissipation occurring in direct contacts between grains and with the walls. Despite the seeming simplicity of the system, the behavior of dry granular material is very rich and extends from solid to gaseous properties depending on the flow regime. In the absence of a unified framework, granular flows are generally divided into three different regimes. (i) At low shear, particles stay in contact and interact frictionally with their neighbours over long periods of time. This "quasi-static" regime of granular flow has been classically studied using modified plasticity models based on a Coulomb friction criterion (Schofield & Wroth 1968; Becker and Lippmann 1977). The response in terms of velocity or solid fraction profiles is independent of the shear rate (Roux & Combe 2002; GDR Midi 2004). Consequently, if the material remains homogeneous, this state only depends on geometric data (shape and size distribution of the grains) and on the inter-granular friction coefficient. (ii) Upon increasing the deformation rate, a viscous-like regime occurs and the material flows more as a liquid (Forterre & Pouliquen 2008). In this intermediate regime, the particles experience multi-contact interactions. (iii) At very high velocity, a transition occurs towards a gaseous regime, in which the particles interact through binary collisions (Goldhirsch 2003; Jenkins and Savage 1983).

For the modelling of dense granular flows, the concept of inertial number $I$ has been widely used and investigated with regard to its relationship with dynamic parameters, such as velocity, stress, and friction coefficient which leads to constitutive relations for granular flows. Thus, '*dynamic dilatancy*' law and '*friction*' law were deduced from discrete simulation of two dimensional simple shear of a granular material without gravity (da Cruz et al. 2005). Those results establish that the flow regime and rheological parameters depend on a single dimensionless number that represents the relative strength of inertia forces with respect to the confining pressure, or the combined effect of pressure and shear rate. Indeed, for a three-dimensional granular medium made up of monodisperse spheres ($d$, $\rho$ resp. the particle's diameter and density) undergoing simple shear flow at a shear rate $\dot{\gamma}$ under an applied normal confining stress $\sigma$, this dimensionless number – inertial number – is defined as $I = \dot{\gamma} d / \sqrt{\sigma/\rho}$. This inertial number also equivalent to the square root of the Savage number or Coulomb number, previously introduced by some authors as the ratio of collisional stress to total stress (Savage 1984; Savage & Hutter 1989; Ancey et al. 1999, see also GDR Midi 2004; Baran & Kondic 2006; Hatano 2007; Luding 2008). This dimensionless number can also be seen as the ratio of $d/\sqrt{\sigma/\rho}$, the time scale for grains to rearrange due to the confining stress $\sigma$, to the time scale $\dot{\gamma}^{-1}$ for deformation by the flow. Hence, it characterizes the local ''rapidity'' of the flow. Thus it was observed that both dimensionless quantities: the internal friction coefficient $\mu = \tau/\sigma$ and the solid fraction $\phi$ are functions of $I$ (GDR Midi 2004; da Cruz et al. 2005; Hatano 2007). Thereby, the inertial number $I$ opened a new path unifying, in a single phenomenological law, many experimental and numerical data in a wide variety of transient flows from the rotating drum to inclined plane flows, where large flowing zones form. A local relation between an apparent friction coefficient and $I$ then successfully captures many aspects of these rapid granular flows (Savage & Hutter 1989; GDR Midi 2004; Jop et al. 2006; Forterre & Pouliquen 2008).

Following general results from simulations of planar shear (da Cruz et al. 2005; Iordanoff & Khonsari 2004), and successful applications to inclined plane flows (Pouliquen & Forterre 2002; Silbert et al. 2003), the experiments of Jop and co-workers (Jop et al. 2006) were carried out to quantify, for glass beads, the $\mu(I)$ – rheology from the quasi-static to the rapid flow regime, corresponding to moderate inertial number (from 0.01 to 0.5) as:

$$\mu = \mu_s + (\mu_2 - \mu_s)/(1 + I_0/I) \qquad (1)$$

in which $\mu_s, \mu_2$ and $I_0$ are three fitting parameters dependent on material properties. According to this law, the internal friction coefficient $\mu$ goes from a minimum value $\mu_s$ for very low $I$ up to an asymptotic value $\mu_2$ when $I$ increases. The asymptotic value of $\mu$ at high inertial number was not obtained by da Cruz and co-workers who observed an approximately linear increase of the internal friction coefficient from the static internal friction value:

$$\mu = \mu_s + aI \qquad (2)$$

where $\mu_s$ and $a$ are two fitting parameters which depend on material properties. In 3D-simulation studies, Hatano did not either observe the asymptotic value of $\mu$ at high $I$ : from $I = 10^{-4}$ to $I = 0.2$, he reported a law in which the friction coefficient increases as a power of the inertial number (Hatano 2007).:

$$\mu = \mu_s + aI^n \qquad (3)$$

It should be pointed out however that this rheology agrees with earlier scaling relations stemming back to Bagnold (Bagnold 1954). Bagnold described a mechanism of momentum transfer between particles in adjacent layers that assumes instantaneous binary collisions between the particles during the flow. Under this assumption, the inverse strain rate is the only relevant time scale in the problem leading, for a constant solid fraction $\phi$, to constitutive relations between the shear stress $\tau$ (resp. normal stress $\sigma$) and shear rate $\dot{\gamma}$ of the form $\tau = \rho d^2 f_1(\phi)\dot{\gamma}^2$ (resp. $\sigma = \rho d^2 f_2(\phi)\dot{\gamma}^2$) where $f_1$ and $f_2$ are functions of the solid fraction $\phi$ only. Bagnold's scaling has been verified for dry grains in both collisional and dense flow regimes (Jenkins & Savage 1983; Lois et al. 1987; Silbert et al. 2001; Lois et al. 2005; da Cruz et al. 2005). The main difference between the *Bagnold* and $\mu(I)$ approaches is that: in the first case, $\phi$ is kept constant for one given flow, whereas it varies freely and depends on the flow in the second case. Indeed, if the pressure is controlled, the solid fraction $\phi$ is free to adapt in the system with the evolution of other parameters. If $\phi$ is fixed however and the normal pressure measured, this will fix the value of $I$. Then when $\phi$ is fixed, the expression of $I$ shows that the normal pressure should scale with the square of the shear rate as it was shown experimentally in annular parallel-plate shear cell (Savage & Sayed 1984).

The applicability of the $\mu(I)$ – rheology has been examined by various studies (Jop et al. 2005, 2006; Hatano 2007; Forterre & Pouliquen 2008; Ruck et al. 2008; Aranson et al. 2008; Peyneau & Roux 2008; Staron et al. 2010; Gaume et al. 2011; Tripathi & Khakhar 2011; Chialvo et al. 2012; Azéma & Radjai, 2014; Gray & Edwards 2014; Edwards & Gray 2015) and many simulations and experiments have shown that the rheology is valid for various flow configurations for different choices of materials, although deviations from this rheology may take place for very slow (quasi-static) flows with small values of inertial number (Aranson et al. 2008; Staron et al. 2010; Gaume et al. 2011). Up to now, there is however in the literature, no experimental data from conventional rheology (*rheometer with conventional geometries*) describing the $\mu(I)$ – rheology of a dry granular material. Indeed, *Couette flows* of granular materials are characterized by the formation of well-defined shear bands that resemble qualitatively the behavior of a yield stress fluid, and make the interpretation of rheometric data tricky. For slow flows, shear banding is generally very strong, with shear bands having a typical thickness of five to ten grain diameters. In this regime of slow flow, the averaged stresses and flow profiles become essentially independent of the flow rate, so that constitutive relations based on relating stresses and strain rates are unlikely to capture the full physics (Schofield & Wroth 1968; Nedderman 1992; Fenistein & van Hecke 2003; Fenistein et al. 2004). Moreover, due to the dilatant behaviour of granular materials, *constant solid fraction $\phi$ experiments* classically made with a Couette shear-cell rheometer are much more difficult to perform than *constant friction coefficient $\mu$ experiments* such as inclined plane flows. When a granular matter is sheared, the spatial distribution of the shear rate is not always homogeneous. Often, shear is localized near the system boundaries with a shear localization width amounting to a few particle diameters. Nevertheless, depending on the boundary conditions, confining pressure and shear velocity, the bulk of the granular system may exhibit different behaviours. For high shear velocities and small confining pressures – *high $I$* –, the granular matter flows homogeneously (Koval et al. 2009). However, at small inertial number, it was indeed shown that in confined annular flow at small shear velocities and high confining pressures – *small $I$* –, the shear may be not homogeneous and solid and fluid phases coexist (Aharonov & Sparks 2002; Jalali et al. 2002).

In the present work, we show that it is not necessary to develop specific set-ups (such as the inclined plane) to study dense granular flows. Indeed, we show that a simple annular shear cell (Carr and Walker 1968, Savage and Sayed 1984) can be adapted to a standard rheometer to study the rheology of granular materials under controlled confining pressure. It allows us in particular to obtain the dilatancy law $\phi(I)$ and also to study very accurately the quasi-static limit. Thus, from the steady state measurements of the torque and the gap during an imposed shear flow under an applied normal confining stress $\sigma$, we report two laws in which the internal friction coefficient and the solid fraction are functions of a single dimensionless number: the inertial number $I$. An effort is then made to compare the present results to the $\mu(I)$ – rheology described in the literature. Indeed, we show that at low inertial number $I$ (small $\dot{\gamma}$ and/or large $\sigma$), the flow goes to the quasi-static limit, and the response in terms of internal friction coefficient $\mu$ and solid fraction $\phi$ profiles is independent of the inertial number. Upon increasing $I$ (large $\dot{\gamma}$ and/or small $\sigma$), $\phi$ decreases while $\mu$ increases. The observed variations are in good agreement with previous observations of the literature. Importantly, we also show that changing the initial gap size does not significantly affect these results. This suggests that shear localization is mostly absent in the intermediate dense flow regime, although it may still occur when the granular material is slowly sheared (in the quasi-static regime).

## II. Materials and methods

To investigate the steady flows of dry granular materials and determine the $\mu(I)$– rheology, three main features are required: (i) to avoid shear banding, (ii) to apply a confining stress in the velocity gradient direction, and (iii) to allow volume fraction variations. If one wants to use a rheometer, (i) implies that the use of a Couette cell should be avoided since it is characterized by a shear stress inhomogeneity that naturally leads to shear-banding. Both the cone-and-plate and the parallel-plate geometries allow a normal force to be imposed in the velocity gradient direction; however, the analysis of the cone-and-plate flow can be performed only at a single gap value, i.e. it cannot be used to characterize a material whose the volume varies under shear. On the other hand, any gap variation in parallel-plate geometry can be accounted for in the determination of the shear rate value. Moreover, shear banding should in principle be avoided in this last geometry since the shear stress is independent of the vertical position in the gap (Macosko 1993). Nevertheless, the shear rate varies along the radial position and is equal to zero in the center. One should thus try to avoid the use of the central zone of the gap. Finally, the material should be confined by lateral walls to make sure that any gap variation actually leads to a material volume fraction variation. These requirements led us to develop a home made annular shear cell, inspired by Boyer et al. 2011, in which pressure-imposed measurements can be performed as shown in Figure 1. Annular shear cells have been extensively used to characterize the flow of pharmaceutical powders and dry granular materials (Carr and Walker 1968, Savage and Sayed 1984, Schulze 1996; Schwedes 2003).

We use a granular material made of rigid polystyrene beads (from Dynoseeds) of density $\rho = 1050\, kg/m^3$, of diameter $d = 0.5$ mm (with a standard deviation of 5%). Spherical beads fill the annular box between two static concentric cylinders with respectively an inner and outer radii of $R_i = 21\, mm$ and $R_o = 45\, mm$. The width of the annular trough is about $48d$ leading to a ratio of inner to outer wall radii of 0.46. We also used another annular channel with the same width but with a larger ratio of inner to outer cylinder radii equal to 0.61 ($R_i = 38\, mm$ and $R_o = 62\, mm$). We have verified that changing the ratio of inner to outer cylinder radii of the annular shear cell does not significantly affect the results. The filling height (initial gap, $h_0$) of the annular box is adjustable from a few grain diameters (typically 5$d$) to 30$d$. The cylinders were finished as smoothly as possible to permit the granular material to slip there as readily as possible. For that, they are made of polyoxymethylene (POM) resin which exhibits a low friction coefficient due to the flexibility of the linear molecular chains. We have measured the friction coefficient at the wall $\mu_w$ between polystyrene beads and a plane made of POM by measuring with our rheometer the sliding stress under different normal stresses such as in Jenike's shear tester (Jenike 1964; Schwedes 2003); it is found to be very small: $\mu_w \approx 0.05$.

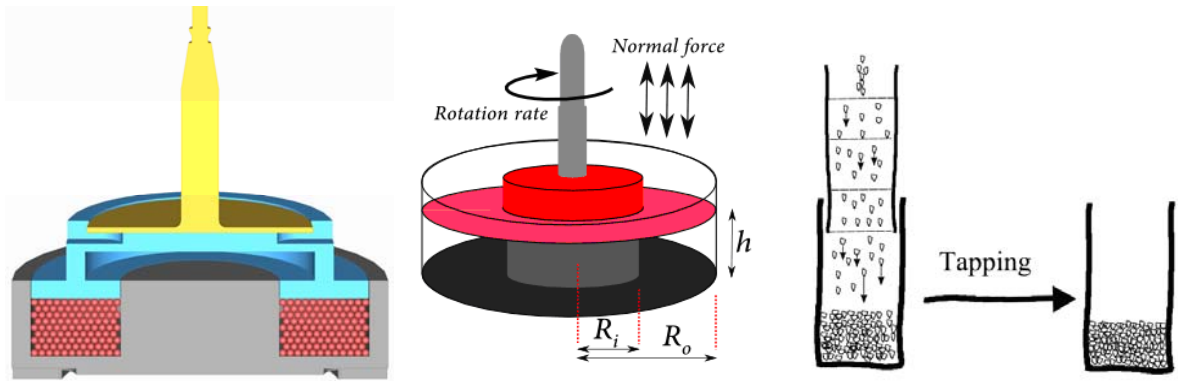

Figure 1: (a-b) Cross section of the annular plane shear flow. The shear and pressure are provided by a ring which is assembled on a *Kinexus* rheometer by *Malvern* that is free to move vertically while maintaining a constant rotation rate or shear rate and imposed pressure. (c) Rain-filling coupled to tapping to get the denser piling sample ('D').

The experiments were performed initially on a very dense piling $\phi_0 \approx 0.625$, close to the so called random close packing of 0.637 (Torquato et al. 2000; Camenen et al. 2012) obtained by combining a rain-filling and tapping the box (Ovarlez et al. 2003) to get a reasonably uniform packing (Figure 1b). However, we will show below that the steady state obtained when the material is sheared is the same for an initially looser piling (Figure 2c). Granular beads are then driven by the ring-shaped upper boundary made of PMMA, which is assembled on a *Kinexus Pro* rheometer by *Malvern*. To avoid wall slip, both the moving upper boundary and the static lower boundary are serrated, with 0.5 *mm* ridges which correspond to the size of grains (Shojaaee et al. 2012).

In our rheometer, instead of setting the value of the gap size for a given experiment, as in previous studies (Schulze 1996; Schwedes 2003) and generally in rheometric measurements, we impose the normal force (i.e. the confining normal stress) and then, under shear, we let the gap size vary in order to maintain the desired value of the normal force. We then have access to instantaneous measurements of the driving torque $T$ and the gap $h$ for imposed normal force $F_N$ and shear rate $\dot{\gamma}$: in this case, the solid fraction $\phi$ is not fixed, but adjusts to the imposed shear. However, it remains important to notice that, in order to keep the imposed shear rate constant, the rheometer adjusts the rotation velocity $\Omega$ since the gap varies as will be discussed below.

A typical measurement is shown in Figure 2, where we start out with a given gap $h_0 \approx 6d$, impose a constant shear rate $\dot{\gamma}$ and normal force $F_N$ and measure the torque $T$ and the gap $h$ as a function of strain (or time). The system reaches a steady state after a certain amount of shear strain but we carefully compare the transient dynamics of these quantities, beginning with freshly poured grains for two preparations, 'rain piling' and 'rain coupled to tapping piling' (resp.) which allow us to get the looser ('L') piling and the denser ('D') piling (resp.) samples with an initial solid fraction of 0.612 and 0.625 (resp.).

Let us start with the imposed quantities. In both cases, a small variation of the rotation velocity is seen (Figure 2a). As discussed above, this variation is a signature of the shear-induced gap variations; it ensures that the shear rate is constant at any time or strain (see inset). We also see that, in both cases, the normal force reaches quickly the stationary targeted value and remains steady during the whole experiment (Figure 2b). Nevertheless, the short transient behavior is different: in the looser sample ('L'), a significant decrease of the normal force is initially observed while a very sharp increase is observed in the densest piling ('D'). The reason is that the densest system initially wants to dilate, and the rheometer response is not fast enough to allow for this fast dilation at constant normal force. By contrast, the loose sample initially wants to compact, and again the rheometer response is not fast enough to allow for this fast compaction at constant normal force.

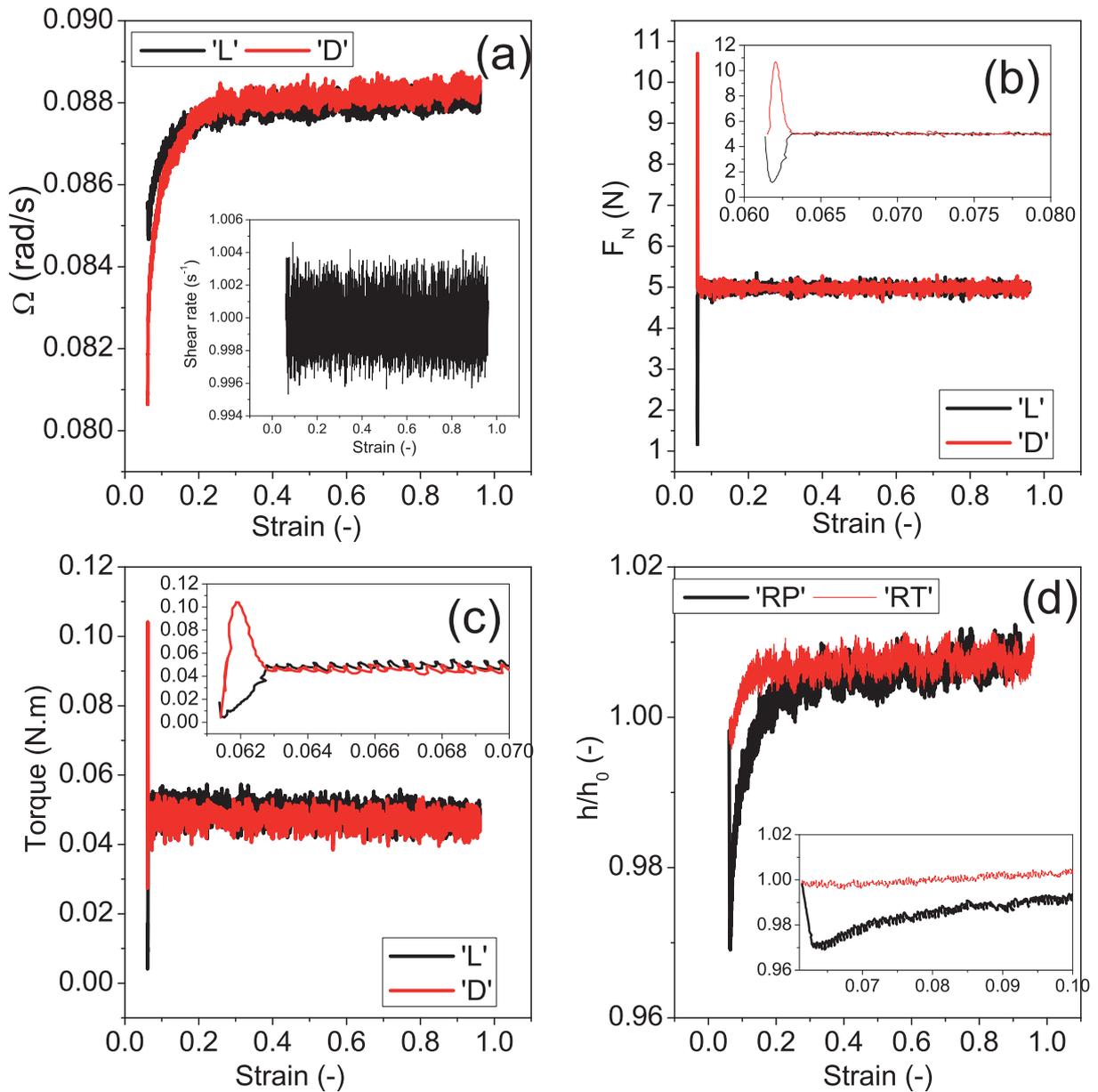

Figure 2: Comparison between samples from the two different pilings: rain piling (which allows us to get the looser piling sample 'L') and rain piling coupled to tapping (that lets us to get the denser piling sample 'D') of the (a) rotational rate against strain; Inset: the imposed shear rate vs. strain, (b) imposed normal force as a function of strain, (c) driving torque and (d) gap size rescaled by the initial gap against strain. The insets in (b-c-d) are the same data showing the sharp increase/decrease of the normal force, the overshoot of the torque and the gap change at the beginning of shearing.

The transient dynamics of the measured quantities (torque and gap size) also shows a clear difference between the two samples. In Figure 2c, we see that, in the loosest piling ('L'), the torque increases before reaching the steady state regime. By contrast, there is an overshoot within the densest sample ('D'), followed by a fast decrease of the torque: the peak strongly depends on the imposed normal force and/or on the shear rate; the decrease corresponds to the continuous dilation of the material under shear. Indeed, in the same time, the gap is not fixed, but adjusts to the imposed shear (Figure 2d). Shear dilation is observed: the measured gap increases up to a constant value since the average of the initial solid fraction is very close to the random close packing. In this regime, shearing necessarily implies immediate dilatancy of the granular media. In contrast, an initially looser sample first compact instead of dilate under shear: when the system is driven under a fixed normal pressure, the granular packing undergoes compaction before dilation into the final state once sufficiently compacted (Wroth 1958).

The most important observation is that the same steady state is reached (same torque and same solid fraction) for the two initial states. This indicates that the material history has been erased, and that we are actually studying the material steady response as a function of $\dot{\gamma}$ and $F_N$ only. As a consequence, there is no need for the sample preparation for the grains we study, namely monodisperse spheres (note that this might not be a general result). All the annular shear cell requires is that the sample is sheared at the required normal pressure until the critical state is reached. However it is necessary to adopt a reference packing state and ensure that the piling method used allows us to achieve sufficient repeatability for a given set materials and system parameters. Hence, we prepare all samples in the same way: a very dense material, from a rain piling coupled to tapping, is used in all the experiments presented below (the initial average volume fraction is 0.625).

## III. Experimental results and analyses

We have measured the driving torque and the gap as a function of shear strain for various imposed normal force $F_N$ (between 1 and and 5$N$) and applied shear rate $\dot{\gamma}$ (between 0.01 and 77 s$^{-1}$) for a given gap $h_0 \approx 6d$.

In Figure 3 (a, b), we show those measurements for $F_N = 3N$ and various $\dot{\gamma}$. At low shear rate, the driving torque increases slowly before reaching a steady plateau within strain of order of unity. Meanwhile, the gap fluctuates around its initial value (we show the gap size rescaled by its initial value before shearing $h/h_0$ called the rescaled gap in the following.).

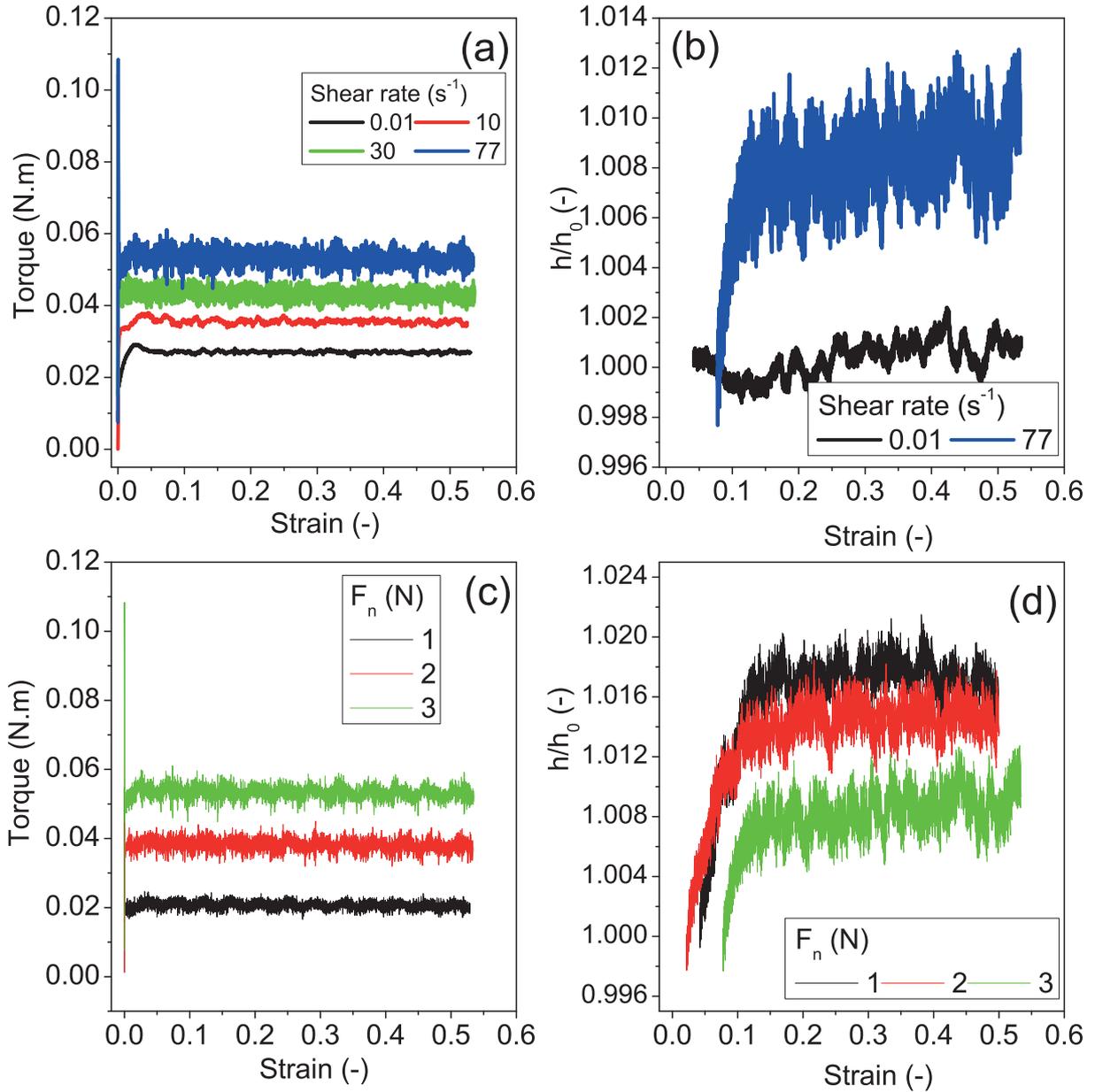

Figure 3: Evolution as a function of the strain at 3N imposed normal force under different applied shear rates of: (a) the driving torque and (b) the rescaled gap size (only 2 curves are shown for clarity). Evolution as a function of the strain at 77 s$^{-1}$ imposed shear rate under different applied normal forces of: (c) the driving torque and (d) the gap size rescaled by its initial value before shearing.

Upon increasing the imposed shear rate, an overshoot occurs: its amplitude increases with increasing the shear rate. In steady state, a rate dependence of the torque is observed (Figure 3a). Moreover, increasing the imposed shear rate causes an increase of the gap size (Figure 3b) allowing to quantify the dynamic dilatancy of the granular material. Notice that, with increasing the applied shear rate, large fluctuations of the driving torque and also of the gap size evolution occur in steady state flows.

Similarly, in experiments in which different normal forces are imposed at a given shear rate (Figure 3c, d), the steady torque is observed to increase while the steady solid fraction (steady gap) decreases when the normal force is increased.

Once the above described experiments are combined, we can obtain the constitutive laws of the dry granular material i.e. the dependence of the steady solid fraction $\phi$ and the ratio between shear and normal stresses $\tau/\sigma$ variation on shear rate. Indeed, from macroscopic quantities $T$, $\Omega$, $F_N$ and $h$, the shear stress $\tau$, the normal stress $\sigma$, shear rate $\dot{\gamma}$, and the solid fraction $\phi$ can be computed.

In the annular plate-cup shear geometry, the driving torque is converted into shear stress using the equation:

$$T = \int_{R_i}^{R_o} 2\pi.\tau.r^2 dr \qquad (4)$$

where $\tau$ is the shear stress, and $R_i$ and $R_o$ are inner and outer radii of the annular trough.

If one neglects the radial velocity gradient (Cleaver et al. 2000; Coste 2004) into the annular trough, the shear stress is quasi-independent of the radial position $r$ and thus, integrating Eq. (4) yields the shear stress as:

$$\tau = 3T/2\pi(R_o^3 - R_i^3) \qquad (5)$$

Eq. (5) holds because the lateral contribution of wall frictions on the stress distribution within the granular media can be neglected. Indeed, for $h = 30d$, this relative contribution can be estimated to be of the order of $\mu_w/\mu$ and we will see below that $\mu_w/\mu \ll 1$.

The normal stress can be also calculated from the normal force as follow:

$$\sigma = F_N/\pi(R_o^2 - R_i^2) \qquad (6)$$

Notice that for $h = 10d$, the imposed normal stress is larger than the hydrostatic pressure $\rho g h$ once $F_N$ is larger than 0.25N, meaning that gravity may be neglected for the range of imposed normal forces.

Besides, assuming that the velocity gradient is approximately uniform over the depth and width of the annular trough and a no-slip condition exists at the rough upper and lower shearing walls, one can estimate the mean shear rate as:

$$\dot{\gamma} = \Omega(R_o + R_i)/2h \qquad (7)$$

And the mean shear strain is given by:

$$\gamma = \theta(R_o + R_i)/2h \qquad (8)$$

where $\theta$ is the angular displacement.

With this analysis, one can plot the shear stress in the steady state as a function of the normal stress as in Figure 4a. The first observation is that a linear relationship between the shear and normal stresses is seen for all imposed shear rates, with a slope that increases from 0.265 to 0.6 with increasing the shear rate. If an internal friction coefficient $\mu$ is defined as the ratio between shear and normal stresses, we evidence here that $\mu$ is rate dependent: it increases with $\dot{\gamma}$.

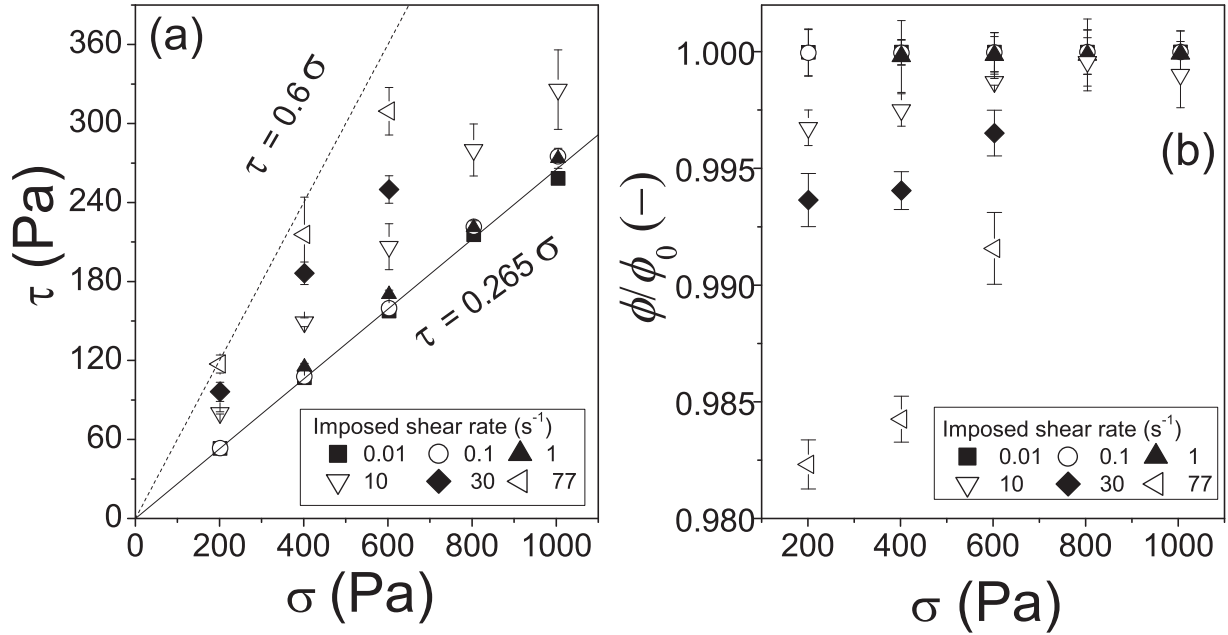

Figure 4: Plot of the shear stress (a) and of the solid fraction rescaled by its initial value before shearing (b) as a function of the normal stress. The stress and the gap are measured in the steady state, for different imposed shear rate. The error bars come from three experiment runs.

The second observation is that, for all imposed shear rates, the steady value of the solid fraction decreases when one decreases the normal stress (Figure 4b). Indeed, since the grains cannot escape from the cell, one can measure unambiguously the solid fraction from the gap variation as:

$$\phi = m / \pi \rho h (R_o^2 - R_i^2) \qquad (9)$$

in which *m* represents the mass of grains. Thus $h_0 / h = \phi / \phi_0$ will simply reflect the impact of shear and confinement on dilatancy.

In order to analyse the results in term of $\mu(I)$– rheology, one has to define an inertial number such as:

$$I = \dot{\gamma} d / \sqrt{\sigma / \rho} \qquad (10)$$

in which $\sigma$ is the normal stress in the steady state regime. It varies between $10^{-7}$ and 0.1 in the range of applied normal force and shear rate. This corresponds to the usual range of quasi-static to dense flow regimes. It should be noted that our annular shear geometry does not allow higher values of $I$ to be studied.

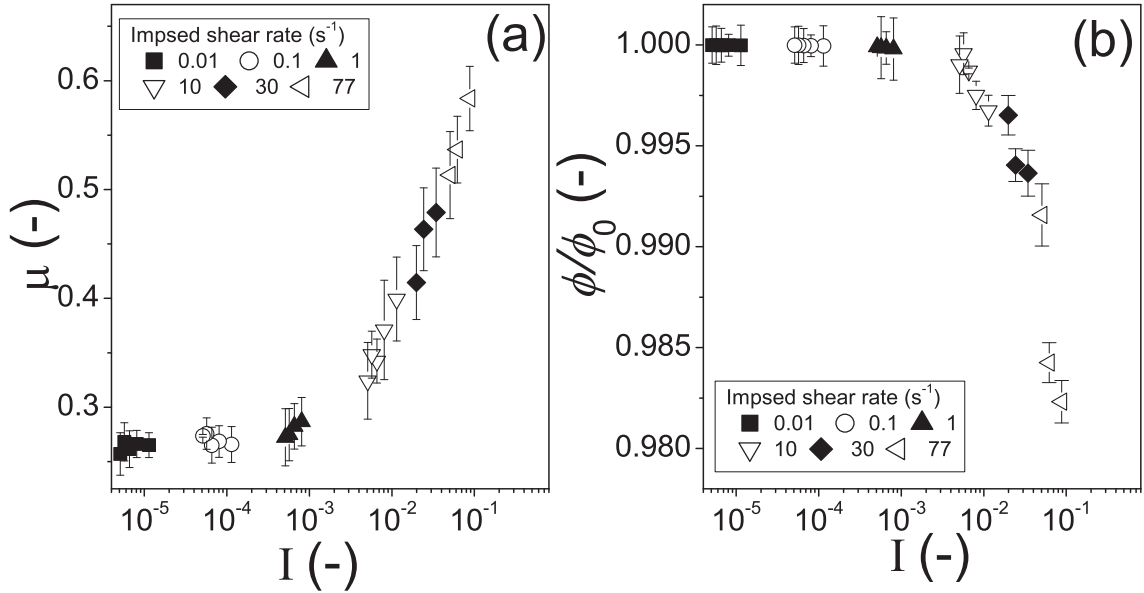

Figure 5: Constitutive law for different sets of mean shear rates and imposed normal forces (a) '*friction law*' i.e. effective internal friction coefficient as a function of inertial number; (b) '*dynamic dilatancy law*' i.e. solid fraction as a function of inertial number. The error bars come from three experiment runs.

Figure 5a shows how $\mu$ vary throughout the flow regimes since $I$ characterizes the local ''rapidity'' of the flow. All the data obtained for different sets of shear rate and normal force collapse on a single curve $\mu = \tau / \sigma$ vs. $I$. For low inertial number, the internal friction coefficient tends to a finite value $\mu_s \approx 0.265$ and increases with increasing $I$. At the same time, the solid fraction variation $\phi/\phi_0$ with the inertial number $I$ is shown in Figure 5b. Once again, all the data collapse on a single curve. At low $I$, $\phi/\phi_0$ is quasi-constant: this is the quasi-static regime. When $I$ increases, the inertia starts influencing the flow and the system becomes rate dependent: the ratio $\phi/\phi_0$ decreases; this regime corresponds to the dense flow in which the granular material dilates.

Moreover, in order to check the robustness of our results, we have varied the initial size of the gap. Here, with our annular shear geometry, the same experiments discussed above are made with different gap sizes from 6$d$ to 22$d$ and different sets of imposed shear rate and normal force (Table 1).

| **Shear rate** (s$^{-1}$) | **Normal force $F_N$** (N) | **Gap** ($h_0/d$) |
|---|---|---|
| 0.01 | 1-2-3-4-5 | 6-8-15 |
| 0.1 | 1-2-3-4-5 | 6-8-10 |
| 1 | 1-2-3-4-5 | 6-10-15 |
| 10 | 1-2-3 | 6-15 |
| 30 | 1-2-3 | 6-15-22 |
| 77 | 1-2-3 | 6-15-22 |

Table 1: Values of the imposed shear rate and normal force for each gap size.

It is shown in Figure 6 that changing the gap does not significantly affect these results. This suggests a total absence of shear localization at high inertial number. However, at small inertial number, the resolution of our measurements is not sufficient to dismiss the possibility that shear localization arises. It was indeed shown that in confined annular flow at small shear velocities and high confining pressures – small –, the shear may be not homogeneous and solid and fluid phases coexist (Aharonov & Sparks 2002; Jalali et al. 2002).

We now compare these experimental results with existing models such as those of Jop et al. Eq. (1) and Hatano Eq. (3) on dry granular flows in terms of quasi-static and dense flow behaviours. Concerning the '*friction law*', our experimental data include points in range of $I$ from $10^{-7}$ to 0.1 which covers quasi-static and dense flow regimes. Our measurements then show that, in this range of $I$, both models can describe our data. Using a classical least squares method, the fit of the data gives: $I_0 \approx 0.032 \pm 0.003$, $\mu_s \approx 0.271 \pm 0.002$ and $\mu_2 \approx 0.68 \pm 0.02$ for the '*Jop et al. model*'; $\mu_s \approx 0.259 \pm 0.002$, $a \approx 1.12 \pm 0.05$ and $n \approx 0.49 \pm 0.02$ for the '*Hatano model*'. In Figure 6b, we show the '*dynamic dilatancy law*', i.e., the variation of the solid fraction $\phi$ as a function of the inertial number *I*. We observe that $\phi$ decreases linearly with *I*, starting from a maximum value $\phi_c$ in the quasi-static regime where the granular material is very dense, close to the maximum solid fraction. The '*dynamic dilatancy law*' is thus given as: $\phi/\phi_c = 1 - (1 - \phi_m/\phi_c)I$ with $\phi_c \approx \phi_0 \approx 0.625$ and $\phi_m \approx 0.495 \pm 0.002$ in agreement with (Pouliquen et al. 2006; Jop et al. 2005).

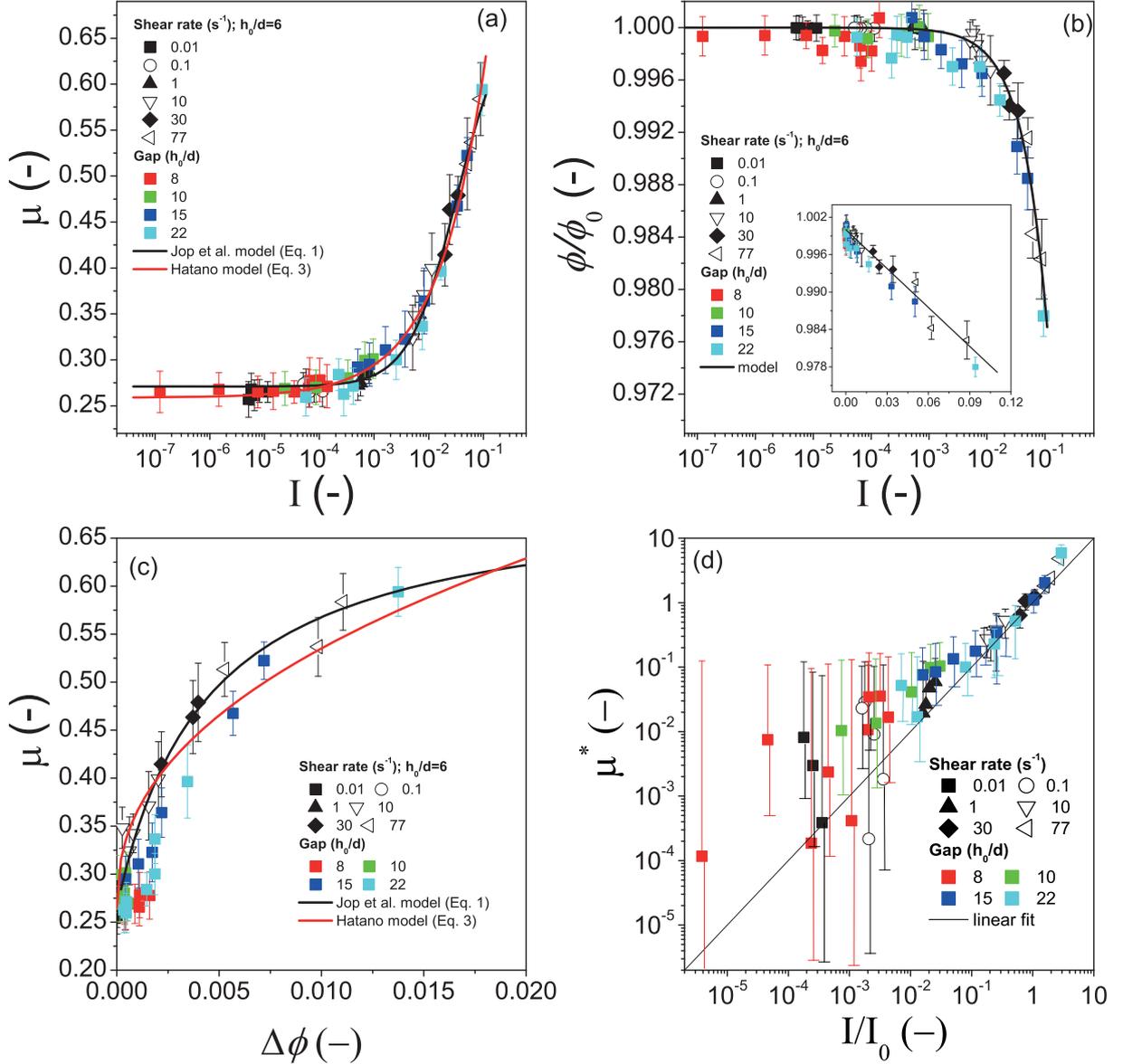

Figure 6: Constitutive law for different sets of mean shear rates and imposed normal forces, and different gap sizes (Table 1): (a) '*friction law*' i.e. internal friction coefficient as a function of inertial number; (b) '*dynamic dilatancy law*' i.e. solid fraction as a function of inertial number; inset: the same data in linear scale. The solid line is $\phi/\phi_c = 1 - (1 - \phi_m/\phi_c)I$ with $\phi_c \approx \phi_0 \approx 0.625$ and $\phi_m \approx 0.495$; (c) Internal friction coefficient vs. the reduced solid

fraction ($\Delta\phi = \phi_c - \phi$); (d) Reduced internal friction coefficient as a function of the reduced inertial number. The line is $\mu^* = I/I_0$. For each gap, the imposed shear rate varies from 0.01 to 77 $s^{-1}$ depending of the imposed normal force in order to obtain either a low $I$ (small $\dot{\gamma}$ and/or large $F_N$) or a high $I$ (large $\dot{\gamma}$ and/or small $F_N$). The error bars come from three experiment runs.

Combining the '*dynamic dilatancy*' and the '*friction*' laws, these data show that the internal friction coefficient $\mu$ strongly depends on the solid fraction: it decreases towards $\mu_s$ when $\phi$ tends to the maximum solid fraction (as shown on Figure 6c in which $\Delta\phi = \phi_0 - \phi$ is the reduced solid fraction). Moreover, following Staron and co-workers (Staron et al. 2010), we defined a reduced internal friction coefficient $\mu^*$ as:

$$\mu^* = (\mu - \mu_s)/(\mu_2 - \mu) \tag{11}$$

We plot in the main panel of Figure 6d the resulting $\mu^*$ vs. $I/I_0$ data points. It holds remarkably well with a prefactor of unity. Satisfying the '*Jop et al. model*' implies indeed that $\mu^* = I/I_0$ which is almost the case for our data except for small values of $I$ ($I < I_0 = 0.03$) wherein deviations seem to take place. The relationship between $\mu^*$ and the reduced inertial number is not clear. These deviations, observed in the quasi-static regime, were also mentioned in recent studies which indicate that this rheology (the '*Jop et al. model*') may be not sufficient to describe the complex phenomena occurring at the flow threshold such as intermittent flows (Mills et al. 2008), and is maybe strictly valid only for relatively large inertial numbers (e.g., $I > 0.02$ as noted by Staron et al. 2010 and $I > 0.005$ by Gaume et al. 2011). The reason for that is not clear yet but one possible explanation is shear localization that arises most often in confined annular flow at small imposed shear and high pressure (small inertial number *I*) (Aharonov & Sparks 2002; Jalali et al. 2002; Koval 2009). Such behavior of granular materials has not yet been fully understood and no consistent and general formalism can predict it successfully (Kamrin 2012). In contrast to discrete numerical simulations (Wang et al. 2012) and theoretical studies (Jenkins & Richman 1985; Richman & Chou 1988; Jenkins 1992), the study of shear localization structure with experimental methods is rather difficult. The visualization of the granular interface is usually limited to the free surface or bottom layers (Fenistein and van Hecke 2003). Recently, MRI has been used to study the granular rheology (velocity and solid fraction profiles) inside the granular system (Moucheront et al. 2010); this tool may thus be a great help in understanding the behaviour at low I. A change in the roughness of a boundary bottom might be used to modify the flow properties such as the wall slip velocity (Shojaaee et al. 2012).

## IV. Conclusion

We have developed a rheometrical method in order to study dense granular flows with a rheometer under imposed confining normal stress and applied shear rate. From the steady state measurement of the torque and the gap, the internal friction coefficient, the solid fraction and the inertial number $I$ are measured. For low $I$, the flow goes to the quasi-static limit and the internal friction coefficient and the solid fraction profiles are independent of $I$. Upon increasing $I$, dilation occurs and the solid fraction decreases linearly when $I$ increases while the friction coefficient increases. The observed variations are in good agreement with previous observations of the literature. As a consequence, we bring evidence that rheometric measurements can be relevant to describe dry granular flows. However, additional experimental work should be carried out in order to measure the dependence of the boundary layer constitutive law on the state of the bulk material, so as to be able to describe properly the rheology when approaching the quasi-static limit.

## Acknowledgments

The authors thank P. Mills for useful discussions, a critical reading of the manuscript and many useful remarks.